\documentclass[letterpaper,twocolumn,10pt]{article}
\usepackage{usenix}

\usepackage{tikz-cd} 
\usepackage{tikz}
\usepackage{adjustbox}

\usepackage{amssymb, amsthm, amsmath}

\usepackage{hyperref}
\usepackage{cleveref}

\usepackage{xcolor} 
\usepackage{listings}
\usepackage{makecell}

\usepackage[frozencache]{minted}

\usepackage{xspace}

\theoremstyle{definition}

\newcommand{\code}[1]{\texttt{#1}}

\newcommand{\latmeet}{\sqcup}

\newcommand{\latorder}{\sqsubseteq}

\crefname{definition}{Defn.}{Defns.}

\crefname{example}{Ex.}{Exs.}

\crefname{figure}{Fig.}{Figs.}

\crefname{section}{Sec.}{Secs.}

\crefname{table}{Tab.}{Tabs.}

\definecolor{mGreen}{rgb}{0,0.6,0}
\definecolor{mGray}{rgb}{0.5,0.5,0.5}
\definecolor{mPurple}{rgb}{0.58,0,0.82}
\definecolor{backgroundColour}{rgb}{0.95,0.95,0.92}

\lstdefinestyle{CStyle}{
    backgroundcolor=\color{backgroundColour},   
    commentstyle=\color{mGreen},
    keywordstyle=\color{magenta},
    numberstyle=\tiny\color{mGray},
    stringstyle=\color{mPurple},
    basicstyle=\footnotesize,
    breakatwhitespace=false,         
    breaklines=true,                 
    captionpos=b,                    
    keepspaces=true,                 
    numbers=left,                    
    numbersep=5pt,                  
    showspaces=false,                
    showstringspaces=false,
    showtabs=false,                  
    tabsize=2,
    language=C
}

\newcommand{\copygif}{\code{gif\_copy\_img\_rect}\xspace}

\newcommand{\ctrlctx}{\gamma}

\newcommand{\handlebeat}{\code{process\_heartbeat}\xspace}

\newcommand{\handlelogin}{\code{login\_handler}\xspace}

\newcommand{\memctx}{\sigma}

\newcommand{\sslwrite}{\code{ssl3\_write}\xspace}

\newcommand{\toolname}{\textsc{Cheesecloth}\xspace}

\newcommand{\words}{\mathsf{Words}}

\definecolor{LightGray}{gray}{0.9}

\usepackage{enumitem}
\setlist[itemize]{nolistsep,topsep=0.7ex,parsep=0.7ex,leftmargin=*}
\setlist[enumerate]{nolistsep,topsep=0.7ex,parsep=0.7ex,leftmargin=*}

\usepackage{etoolbox}
\makeatletter
\patchcmd{\minted@colorbg}{\medskip}{}{}{}
\patchcmd{\endminted@colorbg}{\medskip}{}{}{}
\makeatother

\BeforeBeginEnvironment{minted}{\medskip}
\AfterEndEnvironment{minted}{\medskip\vspace{-1em}}

\usepackage{titlesec}
\titlespacing*{\subsubsection}{0pt}{1.7ex plus 0.4ex minus .2ex}{0.9ex plus .16ex}
\titlespacing*{\paragraph} {0pt}{1.4ex plus 0.3ex minus .2ex}{0.6ex}

\begin{document}

\date{}

\title{\Large \bf Cheesecloth: Zero-Knowledge Proofs of Real-World Vulnerabilities}

\author{
%
%
{\rm Santiago Cuéllar}\thanks{Authors listed alphabetically.}\\
Galois, Inc.
\and
{\rm Bill Harris}\\ 
Galois, Inc.
\and
{\rm James Parker}\\
Galois, Inc.
\and
{\rm Stuart Pernsteiner}\\
Galois, Inc.
\and
{\rm Eran Tromer}\\
Columbia University
} 

\maketitle


\begin{abstract}
  Currently, when a security analyst discovers a vulnerability in
  critical software system, they must navigate a fraught dilemma:
  immediately disclosing the vulnerability to the public could harm 
  the system's users;
  whereas disclosing the vulnerability only to the software's vendor
  lets the vendor disregard or deprioritize the security risk, to the
  detriment of unwittingly-affected users.
  
  A compelling recent line of work aims to resolve this by using Zero
  Knowledge (ZK) protocols that let analysts prove that they know a
  vulnerability in a program, without revealing the details of the
  vulnerability or the inputs that exploit it.
  In principle, this could be achieved by generic ZK techniques.
  In practice, ZK vulnerability proofs to date have been restricted in
  scope and expressibility, due to challenges related to generating
  proof statements that model real-world software at scale and to
  directly formulating violated properties.

  This paper presents \toolname{}, a novel proof-statement compiler,
  which proves practical vulnerabilities in ZK by
  soundly-but-aggressively preprocessing programs on public inputs,
  selectively revealing information about executed control segments,
  and formalizing information leakage using a novel storage-labeling
  scheme.
  \toolname{}'s practicality is demonstrated by generating ZK proofs of
  well-known vulnerabilities in (previous versions of) critical
  software, including the Heartbleed information leakage in OpenSSL
  and a memory vulnerability in the FFmpeg graphics framework.
\end{abstract}



\section{Introduction}
\label{sec:intro}
Ideally, programs that process sensitive information would always
execute safely and securely.
With this ideal remaining difficult to achieve for the foreseeable
future, it is critical that that when programs are found to be
vulnerable, the program's affected users are alerted quickly and
safely.
This requirement presents a challenge: convincingly disclosing
a vulnerability appears to require sharing the vulnerability's details (such as an exploit that triggers it),  thereby placing users at greater risk in the short term.

A promising approach to disclosing vulnerabilities convincingly yet safely
is to leverage \emph{Zero-Knowledge (ZK)
  proofs}: protocols in which one party---designated as the
\emph{prover}---convinces another party---designated as the
\emph{verifier}---of the validity of a claim without revealing any
additional information about the claim's evidence.

Such a use of ZK proofs has arguably been a conceptual possibility
ever since the initial fundamental results establishing that they
exist for all problems in NP~\cite{goldreich1991proofs}.
It has become more realistic with improvements to underlying ZK
protocols and with the emergence of schemes for encoding knowledge of
executions of programs written in convenient languages (starting with \cite{EC:GGPR13,C:BCGTV13} and discussed further below).

In order to prove vulnerabilities in ZK about practical software,
several open problems remaing to be addressed.
First, proof frameworks must scale to compile proofs of
vulnerabilities that require considerably more steps of execution and
space.
TinyRAM~\cite{C:BCGTV13,ben2013tinyram,ben2014succinct} is
sufficiently flexible to validate the executions of applications, but
it is expensive, in part due to the fact that it simulates every
instruction in the modeled CPU's ISA in each step.
TinyRAM's performance is surpassed by those of
Pantry~\cite{braun2013pantry} and Buffet~\cite{wahby2015buffet}, but
both frameworks require loops to be unrolled to a public bound:
publicly revealing these bounds leaks information about the underlying
vulnerability.

A second open problem is to efficiently compile statements from an
understandable form.
One immediate approach is to execute a program under a dynamic safety
monitor for well-understood safety properties, such as those those
implemented in Valgrind~\cite{10.1145/1250734.1250746};
however, directly encoding the additional monitoring would induce
prohibitively large overhead.
Approaches for verifying low-level exploits in
ZK~\cite{green2022efficient} rely on being able to efficiently compile directly-understandable
properties into statements of
control-location reachability.

To address these problems, we present \toolname, an optimizing ZK
proof-statement generator that efficently encodes vulnerabilities in
practical software.
The contributions behind \toolname{}'s design include:

\begin{enumerate}
\item 
  Optimizations of ZK statements that verify the executions of
  programs, taking advantage of program structure but without revealing additional information about the
  execution.
  Specifically, \emph{Public-PC segments} construct execution traces
  from segments with public program counters, thus enabling aggressive
  constant folding, without leaking information about the overall
  execution trace.
  Similarly, instructions which are publicly-determined to be
  executed infrequently are \emph{sparsely} supported (i.e. can't be executed at every step), 
  making the statement smaller.
\item 
  Novel, efficient ZK encodings of memory errors prevalent in practical
  software, specifically \emph{out-of-bound access},
  \emph{use-after-free}, \emph{free-after-free}, and
  \emph{uninitialized access}.
  Previous related work focused primarily on proving knowledge of a
  valid execution without proving existence of a
  vulnerability~\cite{C:BCGTV13,ben2013tinyram} or encoded proofs
  of vulnerability using a less efficient memory
  model~\cite{heath20202}.
\item 
  A novel, efficient encoding of statements that a program \emph{always}
  leaks data (when given an exploit as a secret input).
  Our scheme enables proofs of program properties that are related to, but
  critically distinct from, existing program monitors and type systems
  that prove that a program \emph{may} leak
  data~\cite{enck2014taintdroid,myers1999jflow}, optionally in
  ZK~\cite{fang2021zero}.
\end{enumerate}

We implemented these optimizations and encodings in \toolname, a full
compilation toolchain for encoding vulnerabilities of real-world
programs into efficient ZK proofs.
The toolchain extends previous approaches based on TinyRAM, and
includes a full definition of a novel TinyRAM extension (named
MicroRAM) and a compiler to MicroRAM from the LLVM intermediate
language, enabling proofs of vulnerabilities in programs provided in
C, C++, or Rust.

We evaluated our implementation by proving in ZK the existence of
three vulnerabilities in practical systems software.
Specifically, we proved that previous versions of the GRIT and
FFmpeg~\cite{ffmpeg} graphics processing libraries contained
buffer-overflow vulnerabilites, and that 
the OpenSSL cryptograhy toolkit~\cite{openssl}
%
was vulnerable to the notorious Heartbleed
vulnerability~\cite{heartbleed}.
\toolname{} takes the software C/C++ source code and a flag denoting a
vulnerabilities class; it combines these with an emulation of the
runtime environment (operating system and libraries), and applies the
aforementioned techniques, to derive a statement directly provable in
ZK. The ZK proof can then be given, as a witness, the concrete exploit
used to demonstrate the original attack.
\toolname{} contains implementations of powerful program analyses
that, when combined with manual program partitioning in some cases,
dramatically increase the scale of programs that it can process,
compared to a more naive compiler.

The remainder of this paper is organized as follows:
\Cref{sec:background} reviews the background that this work builds
upon.
\Cref{sec:implementation} presents the implementation details of our
\toolname compilation pipeline;
\Cref{sec:opts} covers the critical and aggressive optimizations we
make to verify the ZK execution of a program;
\Cref{sec:encoding} describes our ZK encodings to efficiently detect
memory and information leakage vulnerabilities;
\cref{sec:eval} describes our practical experience using \toolname{}
to prove vulnerabilities;
\cref{sec:related-work} compares our approach to related work and
\cref{sec:conclusion} concludes.

\section{Background}
\label{sec:background}
In this section, we review prior work on which our contribution builds
upon, specifically \emph{Zero-Knowledge (ZK) proofs} of program executions (\cref{subsec:zk}), %
information leakage by programs (\cref{subsec:bkg-leakage}), and
partial program evaluation (\cref{subsec:partial-eval}).

\subsection{Zero-Knowledge Proofs}
\label{subsec:zk}
\label{subsec:zk-execution}

Zero-knowledge proofs enable a \emph{prover} party to prove to a \emph{verifier} party that the prover knows the correctness of a computational statement (e.g., that a given Boolean circuit is satisfiable), without revealing information about their evidence for the claim (e.g., the witness that satisfied the circuit).
There exist ZK protocols for proving knowledge of
solutions to all problems in NP \cite{goldreich1991proofs}, and
in recent years, numerous efficient protocols have been developed and implemented for ZK proofs of general statements (e.g., \cite{EC:Groth16,EC:GGPR13,SP:PHGR13,C:BCGTV13,ben2014succinct,NDSS:WSRBW15,C:HuMohRos15,EC:MohRosSca17,CCS:AHIV17,AC:BCGJM18,TCC:BHRRS20,heath20202,C:BHRRS21,SP:HYDK21,CCS:FKLOWW21}).

Some of these works specifically address statements about correct execution of programs running on a general-purpose architecture that include Random Access Memory (RAM), where the program is expressed in low-level machine code or a high-level language \cite{C:BCGTV13,braun2013pantry,ben2014succinct,wahby2015buffet,NDSS:WSRBW15,C:HuMohRos15,EC:MohRosSca17,AC:BCGJM18,TCC:BHRRS20,heath20202,C:BMRS21,SP:HYDK21,CCS:FKLOWW21,green2022efficient}.

Our compiler uses a hybrid of step-by-step CPU emulation, similar to
TinyRAM~\cite{C:BCGTV13,ben2013tinyram,ben2014succinct}, a MIPS-like
CPU that can simulate programs in C and similar low-level languages
that access RAM.
The TinyRAM encoder, given a public TinyRAM program and bound on the
number of steps of execution to simulate and a private program input,
generates an R1CS that is satisfied by encodings of the input.
The constraint system consists of %
\textbf{(1)} a family of constraint systems that validate computations
purely over registers in each step and %
\textbf{(2)} a novel memory-checking sub-circuit that verifies the
correctness of RAM operations using a permutation network.
This CPU-unrolling technique is excellent for supporting language
features such as data-dependent loops, control-flow and self-modifying
code. The technique can also naturally leverage existing tools such as
compilers front-ends and libraries.

We combine TinyRAM-style emulation with direct compilation of program
blocks into circuit
gates~\cite{EC:GGPR13,braun2013pantry,wahby2015buffet}
(\Cref{sec:opts}).
The compiler's output is a circuit whose satisfiability is equivalent
to the existence of a vulnerability in the source program, and whose
structure does not reveal the vulnerability or how it may be
triggered.

In our evaluation, the underlying ZK protocol is the Mac'n'Cheese
\cite{C:BMRS21} protocol for proving circuit satisfiability, as
implemented by the Swanky \cite{swanky} library. This is an
interactive protocol, where the prover and verifier engage in multiple
rounds of communication to evaluate the circuit, at the end of which
the verifier learns that the circuit accepted the secret witness
provided by the prover (and nothing else).

\subsection{Information Flow}
\label{subsec:bkg-leakage}
One core contribution of our work is a practical scheme for proving in
zero knowledge that a program leaks data, which we have applied to
prove that previous versions of OpenSSL leak private data, as
triggered by the Heartbleed vulnerability (described in
\cref{subsec:leakage} and \cref{subsec:openssl}).
The scheme's design requires a formal treatment of information flow:
specifically, a treatment sufficiently formal that we could generate
logical circuits that would be satisfied only by witnesses to leakage.
In the interest of space and clarity, we will omit a definition of
information flow and leakage for a full programming language, but we
will describe ours in sfufficient datail to communicate the key
challenges and approaches.

A \emph{labeling} $L$ is a subset of a program's input variables $I$
designated as the \emph{private inputs}, and a subset of its output
variables $O$ denoted as \emph{public outputs}.
Program $P$ satisfies \emph{noninterference} with respect to $L$ if
each pair of inputs that are only distinct at private inputs result in
values that are the same at all public outputs;
$P$ \emph{leaks} with respect to $L$ if, with respect to $L$, it does
not satisfy noninterference.
It follows from the above definition that a leak is witnessed by a
\emph{pair} of executions that differ only at $L$-labeled inputs and
produce distinct $L$-labeled outputs.

Noninterference has a precise but accessible formal definition that
can capture the flow requirements of some critical
software~\cite{enck2014taintdroid}, but its shortcomings in practice
are well
known~\cite{myers1997decentralized,myers1999jflow,sabelfeld2003language}:
the complete information flow specifications of practical programs
often are not noninterference properties, intuitively because
programs that take sensitive inputs typically do need to reveal some
partial information about them;
and even when desired flow properties are noninterference properties,
proving that a program satisfies the property in general can involve
careful reasoning about unbounded data and control.
A rich body of prior
work~\cite{benton2004simple,clarkson2010hyperproperties,denning1976lattice,goguen1982security}
has considered generalizations of noninterference involving
equivalences over observable events, along with rich programming
languages and type systems and attempt to prove their satisfaction.
However, noninterference properties still constitute aspects of a
program's complete information flow requirements that unfortunately
are both critical and are violated in practice (Heartbleed being a prominent example).
This pattern justisfies the current work's primary focus on proving
noninterference violations.

\subsubsection{Labeled Programs and Executions}
\label{subsec:labeled-execs}
In their most general form, information flow and leakage are defined
over pairs of executions.
Practical program
monitors~\cite{enck2014taintdroid,stefan2011flexible} and type
systems~\cite{myers1999jflow} prove facts about all execution pairs, by
\emph{labeling} the program's data and control structures with metadata which is tracked through the execution.
These approaches can be carried out by a programmer or automated
analysis that directly annotates the program or execution.
However, the requisite guarantees are different in our
proof-of-vulnerability context compared to their usual applications,
as seen next.

At a high level, the guarantees provided by dynamic
information flow monitors are as follows. A labeling of a program execution over $n$
steps is an assignment from each program variable and step $0 \leq i < n$
to a sensitivity label.
A labeling \emph{over-approximates} information flow if, from any two
executions starting from states that only differ at high inputs, the
program produces results that differ only at high-labeled timestamped
storage cells (static analyses and type systems lift this property to
be defined over \emph{all} pairs of executions that differ only at sensitive
inputs).

Such over-approximation allows for ``false positives'' in identifying information flows.
For example,
in a context where only input $x$ is
sensitive and the return value is public, the following function
\code{always\_true} does not leak any information about its input
\code{secret} because it returns \code{true} for each input value:
  \begin{minted}[
    bgcolor=LightGray,
    fontsize=\footnotesize,
    ]{C}
bool always_true(bool secret) {
  if (secret) return secret;
  else return !secret; }
\end{minted}
However, many natural taint analyses would label the returned values as
sensitive because it is computed from \code{secret}.

Over-approximation of potential leaks is often still valuable for aiding
programmers to ensure that their program does not leak:
falsely determining that a secure program may leak
may constitute a nuisance, and may need be mitigated to
ensure practicality, but can to some degree to tolerated.

However, in our setting of proving a leak in ZK, it is
unacceptable for the verifier to learn only that a program \emph{may} leak. The whole point is to prove that it \emph{does} leak (given the purported exploit). We will thus create a labeling which is an  \emph{under-approximation}, i.e., when the labels say so, a leakage \emph{is} present. It will then remain to empirically show that labeling indeed detects leakage for the vulnerabilities of interest.

\subsection{Partial Evaluation}
\label{subsec:partial-eval}
In many practical contexts, a program may receive different subsets of
its input at different times after it has been written and compiled:
e.g., after being installed, a configuration file may be included that
remains the same over all executions on distinct inputs subsequently received from
a network.
A natural objective is, given a program and a subset of its inputs that can be fixed, to
generate a specialized program that processes the \emph{remaining} inputs
with improved performance.
Stated more precisely, for program $P(X, Y)$ with input variables $X$
and $Y$, a \emph{partial evaluation} of $P$ on an assignment
$A : X \to \words$ from $X$ to data values $\words$ is a program
$P_{A}$ such that $P(A, B)$ is the same as $P_{A}(B)$ for each
assignment $B : Y \to \words$.

Partially evaluating programs in a practical language brings several
complexities~\cite{jones1993partial};
the underlying technique amounts to:
\textbf{(1)} evaluate the program under a symbolic state, in which
registers and memory addresses may be mapped either to memory
addresses or terms defined over symbolic variables that denoted
unknown values;
\textbf{(2)} using computed symbolic states that describe all possible
states at each control point, simplify the control structure at each
point.
Variations of this technique may be viewed as aggressive extensions
and generalizations of the constant propagation analysis and constant
folding transformation implemented in conventional optmizing
compilers~\cite{aho2007compilers}.

%
%


\section{\toolname Implementation}
\label{sec:implementation}

\toolname produces ZK proofs of real world vulnerabilities.
It takes as input a public LLVM program (typically compiled from C,
C++, or Rust) and, when run as the prover, a secret exploit that
triggers a vulnerability in the program.
\toolname outputs a ZK circuit that verifies the execution trace of
the program and checks whether or not a vulnerability occurred during
that execution.
The pipeline enables a prover to demonstrate to a verifier that there
is a vulnerability in a program while keeping the vulnerability and
triggering exploit secret.

\toolname produces ZK circuits in multiple standard representations
including \emph{Rank-1 Constraints Systems
  (R1CS)}~\cite{C:BCGTV13} and SIEVE IR~\cite{sieve-ir}.
Because the circuits are serialized in standardized formats, \toolname
is agnostic to the ZK protocol applied.
When run as the prover, \toolname outputs the accompanying witness for
the circuit.

\toolname can be extended to check different properties about a
program's execution.
Users can selectively enable which extensions to run by providing different input flags to the compilation pipeline. 
These extensions are how the memory and information leakage vulnerability 
detection checks 
described in \Cref{sec:encoding}
are implemented. 
%
This section covers 
the baseline design of the \toolname compilation pipeline which includes (1) the MicroRAM assembly language, (2) the MicroRAM Compiler, and (3) the Witness Checker Generator. 
\Cref{sec:opts} describes optimizations for this design that enable it to scale to real world vulnerabilities.


%

\subsection{MicroRAM}
\label{sec:microram}

The MicroRAM assembly language is critical to \toolname.
It is the core IR language that \toolname operates on and is the language that the MicroRAM Compiler compiles LLVM programs to.
The Witness Checker Generator produces ZK circuits that verify program executions according to MicroRAM's architecture.

MicroRAM is heavily inspired by TinyRAM \cite{C:BCGTV13,ben2013tinyram}, which
is
a practical and efficient assembly language
with
a simple transition function
that is ideal for ZK execution verification.
We describe MicroRAM and its architecture below, and we precisely describe how its design diverges from TinyRAM in \cref{sec:tinyram}.


MicroRAM is a random-access machine designed to efficiently 
detect vulnerabilities in
program executions.
It is a reduced instruction set computer (RISC) with a Harvard architecture and byte-addressable random-access memory. 

MicroRAM instructions are relatively simple and 
include
4 boolean operations,
8 arithmetic operations for signed and unsigned integers,
2 shift operations,
5 compare operations, 
2 move operations,
3 jump instructions,
2 operations for reading and writing to memory, 
and 
1 answer operation that returns and halts the execution.
Floating-point and vector arithmetic are not directly supported in
the MicroRAM machine
and must be implemented in software.
Instructions take two registers and one operand (either a register or an immediate) as arguments. 
As an example, instruction \texttt{xor ri rj 255} writes to register \texttt{ri} the exclusive-or of register \texttt{rj} and the immediate \texttt{255}.
\toolname extensions like 
those described in \cref{sec:encoding}
can introduce additional instructions as needed.

%

The state of the MicroRAM machine consists of
the program counter (\texttt{pc}),
$k$ 64-bit registers, 
a memory of $2^{64}$ 64-bit words,  
a flag indicating whether or not the execution so far is valid (\texttt{inv\_flag}),
and
a flag 
tracking whether a vulnerability has occurred (\texttt{vuln\_flag}).
\toolname extensions can extend the state of the MicroRAM machine as well.
%
%

To demonstrate the existence of a vulnerability in a program, a prover must present a
secret input that results in a valid execution trace that triggers a vulnerability.
Formally, given a MicroRAM program, \emph{P}, and an initial memory, $m_0$, $P(m_0)$ demonstrates a vulnerability in $T$ steps if
\texttt{inv\_flag} is \texttt{false} and \texttt{vuln\_flag} is \texttt{true}
in
the final MicroRAM state of the program's execution trace. 
\texttt{inv\_flag} is set to \texttt{false} if any of the checks validating the program's execution fails.
% ...
The extensions implementing the vulnerability detection checks
set \texttt{vuln\_flag} to \texttt{true} if they observe a vulnerability during the program's execution.


\subsubsection{Beyond TinyRAM}
\label{sec:tinyram}

As mentioned above, our MicroRAM machine is inspired by TinyRAM. 
Here we report on 
how MicroRAM's design
departs from the TinyRAM model.

\begin{itemize}
  
\item MicroRAM's memory model is byte-addressable while TinyRAM is word-addressable. Byte-addressable memory is necessary to support functionality like string manipulations and packed structs, without adding subroutines to access bytes within full words.

\item TinyRAM receives input via input tapes.
    In MicroRAM, 
    input is passed directly in memory, which saves many cycles that TinyRAM spends copying input to memory. 
    A MicroRAM program can request non-deterministic advice in several ways, 
    however the prover does not have to commit to the advice ahead of time on a tape;
    instead they
    provide the advice upon request.
    This approach is better suited to support backends that exploit
    parallelism or streaming, 
    and it results in smaller circuits.
  
\item TinyRAM uses a 1-bit condition flag for branching 
    while MicroRAM does not. 
    This is advantageous since MicroRAM targets a variety of backends including non-boolean arithmetic circuits where the flag is more expensive than a regular register\footnote{If full words fit in a field element, then the flag is the same size as a register, but requires special circuitry and has more restrictions.}.
    In addition, the semantics without a flag are much simpler
    so the compiler, interpreter, and circuit generator are simpler as well.
    We found that even when targeting boolean circuits, the benefits of having a condition flag are outweighed by the extra complexity.

\item We have not yet explored using a von Neumann architecture \cite{ben2014succinct} for MicroRAM because, despite the asymptotic benefits, the instruction fetching circuit is not yet a limiting factor in our ZK statements.

\end{itemize}

\subsection{MicroRAM Compiler}
\label{sec:clang}

The MicroRAM Compiler is implemented as a LLVM backend that takes LLVM IR programs as input and produces MicroRAM assembly as output.
We currently support C, C++, and Rust programs by compiling them to LLVM IR with the \emph{Clang} and \emph{rustc} compiler frontends. 
Support for other languages such as C\#, Haskell, or Scala can be added in the future by connecting their appropriate LLVM frontends and writing the appropriate standard libraries.

Our compiler backend supports a large subset of the LLVM IR language.
The compiler supports all boolean and arithmetic operations for integers of different sizes, bitwise operations, all non-concurrent memory operations including pointer arithmetic with \texttt{getelementptr}, conversion operations, function calls, variable arguments, comparisons, and \texttt{phi} nodes. 
Complex operations like floating-point operations are 
implemented in software via a LLVM compiler pass. 

Exceptions, and all exception handling instructions, are not supported; but we
we can still tolerate programs with exceptions as long as
the prover is disclose
that the execution of interest, which triggers a vulnerability, does not throw any exceptions.
This is since the MicroRAM Compiler
translates all exception handling instructions to traps that mark the trace as invalid by setting the \texttt{inv\_flag} flag.
By inserting traps, the MicroRAM Compiler can process programs with any number of unsupported features, as long as the prover is willing to reveal that 
those features are not involved in the vulnerable execution.
With this simple trick, users can compile real-world programs without having to 
manually remove unsupported features.
When enabling traps, provers must take care not to reveal too much information about the underlying vulnerability.
\Cref{sec:privacy} presents a more detailed discussion about the security implications of how proof statements can reveal information about their witnesses.
%

\subsubsection{Standard library}
MicroRAM supports a significant portion of the C standard library and 
POSIX 
system calls, using Picolibc~\cite{picolibc}: a library that offers
the C standard library APIs and was originally designed for small
embedded systems with limited RAM.
Picolibc supports multiple widely deployed target architectures,
including ARM, RISC-V, and x86-64.

We implemented MicroRAM as a target architecture for Picolibc.
This enables the MicroRAM compiler to support most of the C standard
library and POSIX system calls.
It is also convenient as it allows prover to publicly customize the
behavior of system calls.
For example, in our case study of OpenSSL, the victim server receives
the malicious request from the attacker over the network.
We customized the behavior of \code{read} when compiled natively to
intercept and record all data received over the network.
When compiled for MicroRAM, \code{read} returns the previously
recorded exploit request, which is loaded from secret memory.
We also customize the implementation of \code{malloc} and \code{free}
to efficiently detect memory vulnerabilities (\cref{sec:dyn-memory}).

\subsubsection{Generating advice}
\label{sec:advicegen} 
As we will see in later sections, \toolname requires nondeterministic advice to 
efficiently generate a ZK circuit that verifies the consistency of memory in an execution (\cref{sec:wcg}) and the presence of a vulnerability (\cref{subsec:mem-encoding}).
To aid the prover in producing that advice, the MicroRAM compiler runs two interpreter passes.
The first pass executes the program without any advice and records the necessary advice.
The second execution runs the nondeterministic semantics and records the trace, which is passed to the Witness Checker Generator to produce the witness for the prover.


\subsubsection{Security}
\label{sec:privacy}

%
MicroRAM produces zero-knowledge proofs which ensure that no \emph{additional information} is revealed about the witness.
However, the proof statement itself can reveal information about the secret input.
For example, in MicroRAM and TinyRAM the
circuit reveals a time bound $T$ on the execution length.
In Pantry/Buffet, the circuit discloses an upper bound $T_i$ on every
loop (and recursive function) in the execution.
In vRAM \cite{zhang2018vram}, every instruction run during the execution is revealed to the verifier.
We argue that a formalization of this information leakage
is necessary.
Interesting and important future work will be to define
a formal framework to analyse how secure these encodings are.

\subsubsection{Preprocessing public inputs}
\label{subsec:public-preproc}
One opportunity for aggressive optimization is to publicly evaluate
logic that is determined by the program's public inputs.
Many practical programs collect inputs from multiple sources, some of which are not secret (i.e., irrelevant to the vulnerability).
If the prover and verifier agree when defining a proof statement that
only some inputs are sensitive secrets (e.g., data packets received
from a network connection) while others are not (e.g., straightforward
configuration options), then the resulting proof statement could be
immediately optimized by generating the proof statement and partially
evaluating the resulting circuit on its input wires corresponding to
non-sensitive inputs.

\toolname supports such cases with a compiler pass that determines the
largest program prefix in which no operation depends on secret inputs.
The MicroRAM compiler then separates the public prefix from the
remaining program suffix and compiles them separately.
When the interpreter is executed by both the prover and verifier, it
executes the prefix and defines a public snapshot of the resulting
state, including both registers and memory.
When executed by the prover, the interpreter then executes the
remaining suffix using both the snapshot and their private input
generate to generate the statement's witness.
In practice, this simple optimization has significant impact, reducing
the number of execution steps 
in OpenSSL's ZK proof statement from
25M to 1.3M (\cref{subsec:openssl}).

The compiler optimization implements a relatively restricted form of
partial evaluation and constant folding (\cref{subsec:partial-eval}).
Our initial experience indicates that further extensions could improve
\toolname{}'s performance drastically: a key technical challenge is
that while programs may perform much processing of public data over
the course of the entire execution, the processing is often
interleaved with computation over sensitive inputs.
Evaluating each of the interleaved phases of public computation is
sound in principle, but can only be automated by ensuring that
regions of storage 
used by public and secret phases are disjoint.
%
Such automation could potentially be achieved by applying points-to
and shape
analyses~\cite{sagiv2002parametric,shapiro1997fast,steensgaard1996points},
including separation logic~\cite{reynolds2002separation}.

\subsection{Witness Checker Generator}
\label{sec:wcg}

The Witness Checker Generator takes as input a MicroRAM program and generates 
a ZK circuit, 
serialized in standardized formats including R1CS and SIEVE IR.
It also accepts nondeterministic advice as input and outputs the secret witness to the circuit when run as the prover.

The Witness Checker Generator builds arithmetic circuits for the prime field $2^{128}-159$.
As an optimization, it automatically constant folds gates that are independent of secret inputs.
To scale to large circuits and avoid running out of memory,
it streams the circuit serialization to a file.
This streaming is independent of secret witnesses, so the same circuit is generated for the prover and verifier.

The nondeterministic advice the Witness Checker Generator accepts provides a description of a program's execution together with the advice necessary to run it.
Concretely, the 
advice
for an execution of $T$ steps contains the initial program memory, 
the $T$ MicroRAM states making up the execution trace,
and
a mapping from step number to additional advice given at each step. 
This additional advice 
includes memory ports for what is read or written to memory and 
stutters that indicate the execution should pause for the current step.


The Witness Checker Generator produces a ZK circuit that
verifies that the witness describes a valid execution trace for the program and that
a vulnerability occurs.
The circuit is split into 
four key pieces: 
(1) the transition function circuit,
(2) the memory consistency circuit,
(3) a state transition network, and
(4) public-pc segments.
We describe the first two here, 
which follow a similar
structure to the circuit
construction for TinyRAM
\cite{C:BCGTV13}. The other two are described later in
\cref{sec:publicpc}.

\paragraph{Transition function circuit.}
The transition function circuit checks a single step of execution.
These checks are chained together to validate the entire execution trace.
\Cref{fig:transfunc} shows pseudocode for the transition function circuit.
It takes as input the circuit's wire representation of the current MicroRAM state, the next state, and any additional advice needed for the current step.
The circuit then fetches the instruction to execute based on the program counter and pulls out the instruction's argument values by indexing into machine registers.
It calculates the expected result of the step by multiplexing over the instruction. 
Finally, the circuit ensures that the calculated expected state matches the next state provided as advice.

\begin{figure}
\centering
  \begin{minted}[
    autogobble=true,
    bgcolor=LightGray,
    fontsize=\footnotesize,
    breaklines=true,
    linenos
    ]{Rust}
  fn transition_func(circuit, current_st, next_st) {
    let expected_st = current_st.clone();
    let instr = fetch_instr(circuit, current_st.pc);
    let arg1 = index(circuit, current_st.regs, instr.op1);
    let arg2 = index(circuit, current_st.regs, instr.op2);

    let result = circuit.mux(instr.opcode == XOR,
                   xor(circuit, arg1, arg2), ...);
    expected_st.pc = circuit.mux(
                       is_jump(circuit, instr.opcode),
                       result, 
                       circuit.add(current_st.pc, 1));
    write_index(circuit, expected_st.regs, instr.dest,
                  result);

    circuit.assert(expected_st == next_st); }
\end{minted}
\vspace{-1em}
\caption{Pseudocode for the transition function circuit that validates a single MicroRAM step.}
\label{fig:transfunc}
\end{figure}

\paragraph{Memory consistency circuit.}
The memory consistency circuit is similar to TinyRAM's except
addresses are byte-addressable instead of word-addressable.
Each step may have a corresponding memory port advice that states the
address and what was read or written to memory.
The transition function circuit verifies that the execution trace
matches the memory port advice.
All of the memory ports are sorted by address and step number.
The memory consistency circuit linearly scans the memory ports to
ensure that all reads and writes to a given address are consistent
with the previous memory operation.
For example, a read should return the same value that was previously
written to an address.
Finally, the memory consistency circuit checks that the sorted memory
ports are a permutation of the memory ports used by the transition
function circuit.
\Cref{subsec:mem-encoding} describes how these checks are enhanced to
efficiently detect memory vulnerabilities.

%
%
%

\section{Optimizations}
\label{sec:opts}
This section describes two of \toolname{}'s key optimizations:
constructing executions with public program counters
(\cref{sec:publicpc}) and tuning steps based on instruction sparsity
(\cref{sec:sparsity}).
\Cref{subsec:opteval} contains an empirical evaluation of the optimizations'
effectiveness.

\subsection{Public-PC segments}
\label{sec:publicpc}

The MicroRAM machine is design to minimize the size of the circuit that checks the transition function circuit.
However, even with MicroRAM's small instruction set, the transition function circuit 
is still large.
%
This is due to the fact that
every transition function must support every operation
in every step of execution. 
What if we could remove \emph{all} the unused functionality? This is
the approach of vRAM \cite{zhang2018vram},
where the circuit is tuned to check 
the instruction that is
executed at each step.
The resulting circuit is much smaller, but unfortunately 
the trace of executed instructions is revealed.
The values in memory and registers would still be kept secret, however a verifier could easily discover where the vulnerable code is in the program.
In this section, we present 
\emph{public-pc segments} which generate
much smaller circuits without revealing the trace. 

A \emph{public-pc segment} is a sequence of transition circuits with a hardcoded and public program counter. Using constant folding, all the instruction fetches of the public-pc segments are known and the unused functionality of every step can be removed.
For example, with a public program counter, the \texttt{fetch\_instr} and subsequent \texttt{mux} operations over the instruction
in \cref{fig:transfunc}
can be constant folded away.
We generate public-pc segments for all straightline code segments in a program, and we implemented a compiler pass that uses a naive control-flow analysis to estimate how many times each segment will be used.
The analysis takes a global bound specifying how many times to unroll loops
and estimates how many times a function will be called by counting the number of call sites for that function in the program. 

To preserve the security of the trace, the cycle counter of segments is kept private.
In addition, we introduce a state routing network
so that the end state of a segment could be the initial state for any other segment in the circuit. 
Just like the memory routing network, the routing information for the state routing network is given by the prover and kept secret.

As a further optimization, we avoid using the state routing network when possible. For example, when a public-pc segment branches to two statically known locations, we directly connect the end state of that segment to the segments representating those two locations.

It is possible that the bound for unrolling loops is not large enough to support certain executions,
so the pipeline would not generate enough public-pc segments for a section of code.  
As backup, the pipeline also produces \emph{private-pc segments} which are just like public-pc segments except the program counter is not revealed.
Private-pc segments look similar to a much smaller TinyRAM circuit with the difference that their start and end states come from the state routing network.
The circuits for these segments are significantly larger, but can execute any part of the program at any point during execution.


\subsection{Sparsity}
\label{sec:sparsity}

With the naive CPU unrolling described in \cref{sec:wcg}, every transition function must contain a memory port,
which causes the memory consistency network to grow at a rate of $O(T \log T)$, where $T$ is the number of steps executed.
Unfortunately, most of those gates are wasted by execution steps that do not
access memory.
\toolname mitigates this excess by removing some of the unused memory ports, thereby
reducing the size of the memory consistency circuit.

The key observation for this optimization is that memory operations
are rarely contiguous.
Even when a program performs a memory-intensive operation, other instructions
are often interleaved between memory instructions.
For example,
when adding the values in a buffer, it takes some steps to increment the
pointer and add the values between 
memory reads.
%
This enables us to share one memory port among $s$ contiguous steps, 
shrinking the memory consistency network by a factor of $s$.

We define the \emph{memory sparsity}, $s$, as the number of steps that share
a single memory port. 
\toolname chooses $s$ based on a static analysis of the
code.
The analysis determines the minimum distance between two memory
operations in any possible execution.
Across statically-unknown jumps
(e.g. calling a function from a pointer dereference), the analysis 
naively considers all the instructions the control flow can possibly
jump to.
This memory sparsity number $s$ is then
used by the MicroRAM Compiler and Witness Checker Generator
to generate the optimized circuit.

Given a memory sparsity $s$, the Witness Checker Generator will group $s$ consecutive steps and create a single memory port for all of these steps.
A multiplexer connects the single memory port to the entire group and sends the result, using nondeterministic advice, to the right step (if any).

If $s$ is larger than the actual sparsity displayed by a trace, then (if unlucky) multiple memory accesses can fall into the same group of steps, which has a single memory port.
\toolname handles this situation
by inserting stutter instructions that delay memory operations until they
are pushed into the next group with separate memory ports.
Inserting stutter instructions can be expensive, but reducing the size of the memory consistency circuit
is more beneficial (\cref{subsec:opteval}).
%
In future work, we will explore swapping program instructions to reduce stutter instructions and determine the optimal $s$ parameter for most programs.



\section{Encoding Vulnerabilities}
\label{sec:encoding}
This section describes how \toolname{} encodes two prevalent and
critical classes of software vulnerabilities: memory unsafety
(\cref{subsec:mem-encoding}) and data leakage (\cref{subsec:leakage}).

\subsection{Memory unsafety}
\label{subsec:mem-encoding}


We now describe how \toolname efficiently models memory and represents
memory vulnerabilties.
In \toolname, memory is an array of $2^{64}$ bytes with half reserved
for the heap and the rest for global variables and the stack.
Our approach is to keep track of valid memory
(e.g. allocated arrays) and report a vulnerability (i.e., set \texttt{bug\_flag}) when the program
accesses non-valid memory.  At the start of the execution, the only
valid memory is where the global variables are stored and, during
execution, \lstinline{malloc} makes allocated regions valid and
\lstinline{free} makes them invalid again. 
With this technique we can catch the following
memory errors:
\begin{itemize}
\item \emph{Uninitialized access}. All uninitialized memory is invalid, so any use triggers a bug.
\item \emph{Use-after-free}. When a region is freed it becomes invalid, so any use triggers a bug.
\item \emph{Free-after-free}. The implementation of \lstinline{free} starts by reading a word from the region to be freed, if the region is not valid it triggers a bug.
\item \emph{Out-of-bound access}. If the program accesses an address out of bounds, that new location \emph{might} (see below) not be valid and this triggers a bug.   
\end{itemize}

It is clear that a normal execution with such bound checking \emph{might} miss out-of-bound access bugs, when the access happens to fall on another valid region, and free-after-free/use-after-free bugs, if an intermediate \lstinline{malloc} makes the region valid before the bug is triggered. However, we only need to show that the bug exists in one execution, so we implement a \lstinline{malloc} guided by nondeterministic advice; this lets the prover choose the allocation layout to ensure the bug is triggered.

While the techniques described here are specific to heap memory bugs, the same ideas can be applied to the stack.

\subsubsection{Encoding dynamic memory allocation}
\label{sec:dyn-memory}

\begin{figure}[t]
  \centering
  \begin{minted}[
    bgcolor=LightGray,
    fontsize=\footnotesize,
    linenos
    ]{C}
void* malloc(size_t size) {
  // Get pointer from advice 
  char* addr = __cc_malloc(size);
  /* Compute and validate the size of
   * the allocation provided by the
   * prover. */
  size_t region_size =
    1ull << ((addr >> 58) & 63);
  /* The allocated region must have
   * space for `size` bytes, plus
   * an additional word for metadata.
   */
  __cc_valid_if(
    region_size >= size + sizeof(uintptr_t),
    "allocated region size is too small");
  /* `region_size` is always a power of
   * two and is at least the word size,
   * so the address must be a multiple
   * of the word size. */
  __cc_valid_if(addr % region_size == 0,
    "allocated address is misaligned for"
    "its region size");
  /* Write 1 (allocated) to the metadata
   * field, and poison it to prevent
   * tampering, invalidating the trace
   * if the metadata word is already
   * poisoned (this happens if the
   * prover tries to return the same
   * region for two separate
   * allocations). */
  uintptr_t* metadata = (uintptr_t*)
    (addr + region_size - sizeof(uintptr_t));
  __cc_write_and_poison(metadata, 1);

  // further computation...
  return (void*)addr; }
\end{minted}
\vspace{-1em}
\caption{Implementation of non-deterministic \code{malloc}. }
  \label{fig:malloc}
\end{figure}

An implementation of \lstinline{malloc} with nondeterminism poses its
own challenges. If left unchecked, the prover could manufacture an
execution that triggers a false bug. For example the prover could
\lstinline{malloc} overlapping regions such that if one is freed and
the other one is accessed a false bug is triggered. Thus,
our implementation of \lstinline{malloc} and \lstinline{free}
(\cref{fig:malloc}) focuses on verifying that the nondeterminisitc
choices are legal. If foul play is detected, the execution is flagged
as invalid with \code{inv\_flag} and will not be accepted by the
verifier.

To ensure that malloc never returns overlapping regions, we
predetermine aligned non-overlapping regions of different sizes for
malloc to choose from. Concretely, we divide memory into $2^6$ pools
of size $2^{58}$, then subdivide pool $i$ into regions of size
$2^i$. \lstinline{malloc} rounds up the requested size to the next
power of two, then returns the start of an unallocated region of that
size.  For example, \lstinline{malloc(15)} must return a region in the
$4$th pool and be 16-byte aligned.  In fact, we can easily verify that
malloc has allocated a correct region just by looking at the pointer
returned: the first 6 bits determine the pool and the rest the
alignment.

Finally, \lstinline{malloc} must not return the same pointer twice
without it being freed in between. To do so, we add to each region one
word reserved for metadata that is marked and made invalid when the
region is allocated. If the region was allocated again, the invalid
metadata would be made invalid again which makes the trace invalid by
setting \code{inv\_flag}.

Furthermore, an implementation of \lstinline{malloc}/\lstinline{free}
that tracks the validity of all memory locations would be quite
inefficient. Luckily, the prover knows exactly where the bug will
happen and thus the \lstinline{malloc}/\lstinline{free} implementation
only needs to track the status of that location. At the beginning of
the execution, the prover commits to a secret location stored in the
global variable \lstinline{__cc_memory_error_address} and then
\lstinline{malloc}/\lstinline{free} only track the validity of that
location. In particular, if an allocated/freed region does not contain
\lstinline{__cc_memory_error_address} then
\lstinline{malloc}/\lstinline{free} do not check for errors, and run
in constant time.

\subsection{Data leakage}
\label{subsec:leakage}
A straightforward approach to proving leakage would be to directly
encode the definition of noninterference in the ZK circuit.
This could be accomplished by verifying two program executions where
only sensitive inputs are distinct but public outputs are distinct.
However, such an approach would result in a statement of twice the
size required for validating a single execution.
Instead, we might hope to prove a leak using a single execution in
which storage is annotated with labels (\cref{subsec:bkg-leakage}).
However, such systems tranditionally have only been designed to prove
that a program \emph{may} leak information, which is unacceptable for
definitively proving a leak without providing a violating execution
directly (\cref{subsec:labeled-execs}).


\paragraph{Specifying leakage}
To identify sensitive sources and sinks, the instructions
\code{source} and \code{sink} are added to the MicroRAM instruction
set, and are directly wrapped by user-level functions
\code{taintSource} and \code{taintSink}, respectively.
\code{source} annotates that a given byte of data carries sensitive
data;
\code{sink} annotates that a given byte is output to a channel. 
Instantiating the general definitions of information flow and leakage
(\cref{subsec:bkg-leakage}) for this extended ISA, a MicroRAM program
leaks if it has two executions whose inputs only differ at addresses
given to \code{source}, but result in different values at an address
given in calls to \code{sink}.
Leakage is established by the prover and verifier collaborating to
extend the subject program to call the \code{taintSource} and
\code{taintSink} to annotate sensitive sources and sinks.


\paragraph{Proving leakage}


To soundly and precisely prove leakage, we propose a novel labeling
system that tracks what program storage may and \emph{must} hold
secret. 
There are four labels, denoted and partially ordered as
\[ \bot \latorder \ell_0, \ell_1 \latorder \top
\]
with a least-upper bound (i.e., \emph{join}) operation denoted
$\latmeet$.
Labels $\ell_0$ and $\ell_1$ annotates data that \emph{must}
belong to one of two principals;
$\top$ denotes that the data's sensitivity is unknown;
$\bot$ denotes data that \emph{must} not be influenced by a
principal.
%
With this labeling scheme, leakage
of $\ell$-labeled data written to a $\ell_c$-labeled sink
 \emph{must} occur when $\ell \not= \top \land \ell \not\sqsubseteq \ell_c$.


MicroRAM state is extended so that every register and byte of memory is associated with a
label, similar to previous leakage
monitors~\cite{enck2014taintdroid,stefan2011flexible,suh2004secure,10.1145/3290388}.
Two additional labels model effects of instructions other than
register arithmetic.
The \emph{control context} label $\ctrlctx$ is maintained to be $\bot$
if the program execution has not branched on secret data, and $\top$
otherwise;
similarly, the \emph{storage context} label $\memctx$ is maintained to
be $\bot$ if the program has not stored to a secret address, and
$\top$ otherwise.

Each assignment \code {x:=e} sets the label of \code{x} to
$L(e) \latmeet \ctrlctx \latmeet \memctx$ (where $L(e)$ is the label of $e$, defined
below);
thus, if the program has branched on secret data or written to a secret address, the label of
\code{x} is set to $\top$.
If $e$ is an arithmetic/logical operation $f(y)$, then $L(e)$ is
$\bot$ when $L(\code{y})$ is $\bot$ and $\top$ otherwise;
%
$L(e) = L(\code{y})$ if $f$ is a
\emph{bijection}: our current implementation conservatively only
labels single-register expressions (i.e., copy sources) as
$L(\code{y})$.
If $e$ is a load \code{*p}, then $L(e)$ is
$L(\code{*p})$. 
%
Conditional branches update $\ctrlctx$ and memory stores update
$\memctx$ according to the labels' descriptions;
we omit formal descriptions here, due to space constraints.

Plenty of natural programs leak but cannot be proved to do so by this
labeling system, potentially because a leakage happens after branching
or storing to a $\top$-labeled value, or because a secret value is
propagated over an operation not recognized as a bijection.
Such cases restrict the situations in which the labeling scheme can be
applied to prove leakage, but do not threaten its validity when it
claims that a given program leaks.
They might be addressed in future work that refines instruction
interpretations using valid logical axioms (e.g., the fact that for
each value $x$, $x + 0 = x$).
Such threats and mitigations are dual to threats to precision, and
possible refinements when proving that a program execution \emph{may}
leak.

\section{Evaluation}
\label{sec:eval}

We evaluate \toolname with three case studies that demonstrate ZK
proofs of real world software vulnerabilities.  The vulnerabilities
scale by code size and execution trace length to showcase the
capabilities of \toolname.  We also benchmark the optimizations
(\cref{sec:opts}) to evaluate their effectiveness.

\Cref{table:casestudies} presents the results of using \toolname to produce ZK proofs for our case studies which include GRIT, FFMPEG, and OpenSSL.
For each case study, we report the size of the program in terms of the number of MicroRAM instructions,
the number of execution steps required to demonstrate the vulnerability, and
the number of multiplication gates in 
the resulting ZK circuit.
We prove satisfiability of the ZK circuit using the Mac'n'Cheese \cite{C:BMRS21} interactive ZK protocol, as implemented by the Swanky \cite{swanky} library.
We record the protocol running time and communication cost between the prover and verifier.
All measurements were performed on a 128 core Intel Xeon E7-8867 CPU
with 2 TB of RAM running Debian 11, although our implementation
typically uses considerably less memory (\cref{table:casestudies}).
%

\begin{table*}[t]
 \centering
 \small
\begin{tabular}{lrrrrr}
\hline
\multicolumn{1}{c}{\textbf{Program}} & \multicolumn{1}{c}{\textbf{Code size (K instrs)}} & \multicolumn{1}{c}{\textbf{Execution steps (K)}} & \multicolumn{1}{c}{\textbf{Mult gates (M)}} & \multicolumn{1}{c}{\textbf{Protocol time}} & \multicolumn{1}{c}{\textbf{Protocol memory}} \\ \hline 
  \multicolumn{1}{l}{GRIT}            & 3                                                 & 5                                                & 26.7                                        & 3m 40s                                    & 845 MB                                                                                   \\
  \multicolumn{1}{l}{FFmpeg}          & 24                                                & 79                                               & 672.7                                         & 1h 22m                                    & 19 GB \\
  \multicolumn{1}{l}{OpenSSL}          & 340                                                & 1,300                                               & 17,049.5                                         & 36h 45m                                 & 460 GB \\
\hline
\end{tabular}
    \caption{Results for generating and running a ZK proof of software vulnerability for each case study.} 

\label{table:casestudies}
\end{table*}

\subsection{Memory unsafety in GRIT}
\label{sec:grit}
The GBA Raster Image Transmogrifier (GRIT)~\cite{grit} converts
bitmap image
files to a graphics format that is readable by the Game Boy Advance. 
%
A bitmap image includes headers, a palette array indicating the colors in the image, and the pixels for the image.
For 24bpp images, GRIT's parser assumes the palette size is zero and 
allocates a 
buffer without space for the palette.
When populating the buffer, it checks the image header for the number of palette entries
without checking that this matches 
%
the assumed palette size that was used during allocation. 
As a result,
a malformed 24bpp image 
can write an arbitrary amount of data (up to the length of the file)
past the allocated buffer.

To demonstrate this memory error, we
construct a 24bpp exploit image with 0x3000 bytes of pixel data and 12 bytes of palette data.
%
%
On Linux, the 12 byte overflow overwrites heap metadata and triggers an
assertion failure in the memory allocator.
When run through \toolname, we generate a ZK proof that 
a memory error is triggered within six thousand steps of GRIT's execution without revealing the triggering image or where the error occurred in the code.
%

\subsection{Memory unsafety in FFmpeg}
\label{sec:ffmpeg}

FFmpeg is a tool for recording, converting, and streaming audio and
video~\cite{ffmpeg}, and is used in popular software projects such as
Chrome, Firefox, iTunes, VLC, and YouTube.
FFmpeg is written in C and has been plagued by vulnerabilities that
compromise memory safety, enabling attackers to execute code and share
local files over the network.
Versions of FFmpeg prior to v1.1.2 contained a
vulnerability~\cite{cve-2013-0864} caused by the memory error in the
function \copygif, which copies the frame of a GIF file between
buffers.
Previous versions of \copygif insecurely calculated a pointer to the
end of a memory buffer by directly using the input image's height.
This calculation allowed an attacker to provide a carefully crafted
GIF which causes FFmpeg to write to memory outside of an array's
bounds.

To prove memory unsafety of FFmpeg in ZK, we manually crafted a GIF
image that exploits the described memory vulnerability.
We passed this image and a program module that invokes FFmpeg's video decoder
%
to \toolname, which generated a proof of a out-of-bound access.
The only facts revealed about the exploitative GIF are what are
implied by the fact that it trigger an out-of-bound access within 76K
steps of execution.
%

\paragraph{Preprocessing FFmpeg on public inputs}
There was potential to aggressively optimize FFmpeg's proof statement,
which was ultimately achieved by applying \toolname{}'s constant
folding transformation pass after manual program partitioning.
The need for partitioning arose due to the interleaving of public and
secret computation in the GIF modules, which executes by:
\textbf{(1)} demultiplexing a given \emph{secret} GIF file into a
sequence of data packets;
\textbf{(2)} initializing the state of the decoder, using
\emph{public} configuration settings;
\textbf{(3)} executing the codec that contains the vulnerability.

Although phase (2) computes entirely over public data, it would not be
optimized by 
\toolname{}'s constant
folding pass because the pass halts upon detecting computation that
uses secret data, and thus would not optimize any program segment
after phase (1).
To address this issue, we manually partitioned the program by phase,
applied \toolname{}'s constant folding pass to each, and linked the
resulting optimized MicroRAM code.
%
%
%
In general, our case study of FFmpeg motivates the further study and
design of more aggressive constant folding passes, which might apply
more sophisticated static program analyses
(\cref{subsec:partial-eval,subsec:public-preproc}).

\subsection{Leakage in OpenSSL}
\label{subsec:openssl}

OpenSSL~\cite{openssl} is a widely deployed open-source cryptographic
library that contains implementations of the SSL and TLS protocols.
OpenSSL versions 1.0.1 to 1.0.1f contained a devastating
vulnerability dubbed \emph{Heartbleed}, discovered in 2014~\cite{heartbleed}, that could be
exploited by a remote attacker to completely leak information stored
over the protocol's execution, including other clients' sensitive
information and private keys.

Comprehensive descriptions of SSL and OpenSSL are beyond the scope of
this paper;
for the purposes of our work, it suffices to note that SSL parties
support multiple requests, including both requests to store data from
the another party and requests to reply to a \emph{heartbeat} signal:
a signal sent only to obtain a response to ensure that the other party
is still responsive.

\begin{figure}[t]
  \begin{minted}[
    bgcolor=LightGray,
    fontsize=\footnotesize,
    linenos
    ]{C}
void process_heartbeat(SSLRequest *req) {
  unsigned int len = parse_heartbeat_len(req);
  unsigned char *heartbeat = get_heartbeat(req);
  unsigned char *response = malloc(len);
  memcpy(response, heartbeat, len);
  write(response, len); }
\end{minted}
\vspace{-1em}
\caption{Pseudocode depicting the Heartbleed vulnerability.}
\label{fig:parse-hb-req}
\end{figure}
The heartbeat request and response is critical to the operation of the
Heartbleed vulnerability.
A well-formed request consists of a \emph{data buffer} $d$ and a
\emph{length field} $n < |d|$.
A correct response to such a request returns the first $n$ bytes
contained in $d$.
However, a party could potentially transmit an \emph{ill}-formed
request, in which $n > |d|$.
The correct response to such an ill-formed request is to reject it.

The implementation of OpenSSL (illustrated by the pseudocode function
\handlebeat in \cref{fig:parse-hb-req}) crucially failed to implement
this aspect of the protocol and instead returned the $n$ bytes of
memory contiguous with the input buffer.
\handlebeat takes a heartbeat request from a client and echos the
provided heartbeat string back.
It does so by first parsing
the length of the heartbeat string from the client's request.
The function then gets a pointer to the heartbeat string in the
request.
Next, it allocates a response buffer and copies \code{len} bytes from
the heartbeat string into the response buffer, which is subsequently
sent back to the client.
Since \code{process\_heartbeat} does not check the provided heartbeat length against the actual length of the heartbeat string,
if the claimed length is larger than the actual length of the
provided heartbeat string, memory beyond the client's request is sent
back to the client. This is practically exploitable, and has been demonstrated to reveal sensitive in-memory data such as cryptographic keys and passwords.

Using \toolname, we proved in ZK that OpenSSL version
1.0.1f
%
%
leaks arbitrary user information in 1.3M steps of execution,
propagating data purely over register copies, loads, and stores;
while the statement reveals a bound on the amount of computation
required to perform the leak and information about the types of
instructions used to perform the leak (described below), it gives no
direct indication of what validation is missing in the function for
processing heartbeat requests, or that heartbeat requests are involved
in the leak at all.
We describe the statement proved, along with technical challenges and
solutions, in more detail below.

\paragraph{Specifying OpenSSL's leakage} A primary challenge of our
work was to provide a scheme for identifying sensitive sources and
sinks such that:
\begin{enumerate}
\item
  A verifier with only an understanding of the data that a
  subject program handles should be able to inspect the modified
  program and definitively conclude that it correctly defines
  information sources and sinks.
\item
  Any modifications to the program to enable the definition of sources
  and sinks between which information is leaked must not reveal
  additional information about the leak's triggering input.
\end{enumerate}

\begin{figure}[t]
  \begin{center}
  \begin{minted}[
    bgcolor=LightGray,
    fontsize=\footnotesize,
    linenos
    ]{C}
int login_handler(
  SSLConn *c, char *password, int len) {
  ...
  label l = getLabel(c);
  for (size_t i = 0; i < len; i++)
    taintSource(password + i, l);
  ... }
int ssl3_write(
  SSLConn *c, char *buf, int len) {
  ...
  label l = getLabel(c);
  for (size_t i = 0; i < len; i++)
    taintSink(buf + i, l);
  ... }
\end{minted}
\end{center}
\vspace{-1em}
\caption{Versions of the OpenSSL functions \code{login\_handler} and \code{ssl3\_write} that we augmented with operations that specify information sources and sinks.
  Passwords are tainted with the label of the current connection, and
  leaks are detected if data written to the network has a label from a
  different connection.}
\label{fig:hb-labelers}
\end{figure}

Our mechanism for defining sensitivity of sources and sinks consists
of the designated functions \code{taintSource} and \code{taintSink} 
(\cref{subsec:leakage}).
We found that such a library served well for specifying information
flow in in OpenSSL;
psuedocode of C functions modified in the OpenSSL codebase to
label sources and sinks are given in \cref{fig:hb-labelers}.
The function \handlelogin, given an SSL connection \code{c}
and a buffer \code{password} presumed to contain \code{len} bytes of
sensitive information to be transmitted over \code{c}, labels
\code{len} addresses beginning with \code{password} with the label of
\code{c}.
The function \sslwrite, given an SSL connection \code{c} and buffer
\code{buf} presumed to output \code{len} bytes, denotes 
sinks at the output channel with the label of \code{c}
for
\code{len}
addresses beginning with \code{buf}. 
%

The modifications to \handlelogin and \sslwrite illustrate the
utility of first-order labels that can be operationally collected and
set, as opposed to operations that set addresses as only high sources
or low sinks, even in a setting in which the information belonging to
only one principal is of interest.
By using first-order labels, we we able to write small specification
functions that only unified the labels between a network connection
and a given buffer, and then succinctly modified the original program
logic in contexts that readily provided a connection and related
buffer.

%

\paragraph{Proving OpenSSL's leakage}
Once OpenSSL has been suitably modified to call the
\code{taintSource} and \code{taintSink} functions, its leakage can be
proved by generating a statement whose solution corresponds
to an execution of a server running OpenSSL that leaks
sensitive data from one connection to another connection.
%
%
We have generated such a statement where the server first responds to a public login request
where the password is marked as sensitive. 
The server then handles a secret malicious heartbeat request
that returns the password from the previous request's connection.
%

Using \toolname{} we prove OpenSSL's leakage
by validating the previously described execution 
which is derived
from one of its
originally disclosed exploits.
The leakage is detected through the source and sink annotations according to our proposed
must-leak labeling scheme (\cref{subsec:leakage}) and the verifier only learns an upper bound on the length of the malicious request.
We found that the labeling scheme enabled leakage to be proved much
more efficiently, reducing the overall circuit size by 30.6\% over the two trace approach.
\toolname{} proved the vulnerability of OpenSSL in approximately 37
hours, using 460 GB of protocol communication.
%

\subsection{Optimizations}
\label{subsec:opteval}

\begin{table}[t]
 \centering
 \small
\begin{tabular}{lrr}
\hline
    \multicolumn{1}{l}{\thead[l]{Program\\ }} & \multicolumn{1}{l}{\thead[l]{Mult gates without\\ public-pc segments (M)}} & \multicolumn{1}{l}{\thead[l]{Mult gates without\\ sparsity (M)}} \\ \hline
    \multicolumn{1}{l}{GRIT}            & 42.3 (37\%) & 27.9 (4\%) \\
    \multicolumn{1}{l}{FFmpeg}            & 716.9 (6\%) & 709.9 (5\%) \\

\hline
\end{tabular}
\caption{The number of 
  multiplication gates in the
  circuit with the different optimizations disabled, as well as the
  percentage increase in size
  over the baseline numbers from \cref{table:casestudies}.
} 
\label{table:opteval}
\end{table}

\Cref{table:opteval} contains the improvements yielded by our key
optimizations (\cref{sec:opts}).
We ran the GRIT and FFmpeg case studies with each optimization disabled and report on
the number of multiplication gates in the resulting ZK
circuit.
In addition, we provide the percentage improvement over the baseline
numbers from \cref{table:casestudies}.
The public-pc optimization reduced gate size by 37\% in the shorter GRIT execution
and 6\% for FFmpeg.
While this is an improvement, these results indicate there is still room for improvement in our analysis that determines
the number of public segments to generate for longer executions. 
The sparsity optimization with $s=2$ offers modest improvements of 4\%--5\% in gate size.

%


\section{Related Work}
\label{sec:related-work}


Recent work has provided the first, exciting steps toward proofs of
vulnerability in ZK.
\code{BubbleRAM}~\cite{heath20202} is an efficient framework for
proving vulnerability, leverage novel protocols for converting between
computations in arithmetic and Boolean fields, efficiently handling
both read-only and read-write memory, and the \code{Stack}
protocol~\cite{heath2020stacked} for proving satisfaction of circuits
with explicit disjunctions.
Although our current statement compiler partially overlaps with
\code{BubbleRAM} because it implements an older scheme for modeling
RAM computations~\cite{C:BCGTV13}, most of our paper's key
contributions, namely simplifying unrolled computations using partial
evaluation and the novel scheme for generating statements of
application leakage, are largely independent of the contributions of \cite{heath20202,heath2020stacked}, and we believe that the approaches
could be composed.
In particular, \code{Stack} was evaluated on code snippets representative of a
practical CVE of up to 50 LoC;
due to its efficient support of disjunctions, it could scale to prove
that one of many more such snippets is vulnerable, but it likely
strongly benefit from \toolname's program optimizations if any
particular code segment increased in size.

Reverie~\cite{green2022efficient} is a framework for proving exploits
in microprocessor code, consisting of a circuit generator that
compiles a given program to an arithmetic circuit and an
instantiation~\cite{katz2018improved} of the ``MPC in the Head''
protocol~\cite{ishai2007zero}.
The compiler generates statements from exploits that have been
formalized as executing a designated instruction that signals an error
condition (i.e., violations of reachability properties, formalized
directly in the program's control flow);
the evaluation of Reverie demonstrates that it can be used to prove
\emph{Capture the Flag} (CTF) exploits that require up to 51K cycles
on an MSP430 microprocessor.
The core contributions of Reverie are largely complementary to those
of \toolname{}, which could potentially be adapted to efficiently
compile vulnerability statements about programs in intermediate
languages to control reachability properties.

Recent work on static program analysis in ZK~\cite{fang2021zero} has
presented techniques for proving over-approximations of all program
executions without revealing further details of the program, and
instantiates the framework on an abstract domain for information flow
based on taint tracking.
The static analysis itself is designed to prove that a program may
leak information: thus, it cannot yield results that directly imply
that a program must leak, although in many cases it could provide
evidence that could strongly inform an analysts belieft that a program
may in fact leak.

Our MicroRAM machine is inspired by TinyRAM~\cite{C:BCGTV13} but
departs from their design in sevaral important ways discussed in
\cref{sec:microram}.
There are also some key differences in scope and
capabilities. TinyRAM is designed to express correctness of any
nondeterministic computations while MicroRAM focuses on vulnerable
programs. For example, \emph{SNARKs for C}\cite{C:BCGTV13}
approach cannot encode proofs of memory-safety vulnerability in ZK
directly. Instead, they encode knowledge of the existence of a
complete, concrete vulnerability trace, which includes copies of exact
values in all local variables and the values in memory at each point
in the trace and the bug must be evident in the execution's return
value. Our approach encodes memory vulnerabilities directly, resulting
in a significantly more succinct witnesses to vulnerability. In
particular we can disregard the trace after the bug is found and we
don't rely on the programs return value.

Furthermore, TinyRAM approach does not scale to proofs of
vulnerabilities in practical programs and has only been evaluated on
programs with less than 1,200 low-level instructions
\cite{C:BCGTV13}.
In contrast, the optimizations proposed in this
work enables us to support programs with more than
340,000
lines of low-level code (\cref{table:casestudies}).
Beyond scalability, MicroRAM supports a much broader subset of the C
language, including most of the standard C library.

Pantry and Buffet \cite{braun2013pantry,wahby2015buffet} represent
computation as arithmetic constraints;
a solution to the constraints is a valid trace of the
computation.
After implementing the memory consistency approach of
TinyRAM, they report results orders of magnitude better than
TinyRAM.
Buffet supports all features in the C language, with the exceptions of
\code{goto} statements and function pointers.
To translate computation into a constraint system, Pantry and Buffet
must unroll loops to publicly revealed bound (although the original
work does not explicitly discuss encoding recursive functions, we
hypothesize that they would be encoded similarly, using bounded
function inlining).
The constraint system must include every branch of conditionals and
every iteration of every loop (multiplicatively with nested loops)
which could lead to blowups in the constraint system, however the
authors suggest that this would only happen in degenerated cases and
would not be common in practice.
A variant Pantry/Buffet that uses zero-knowledge techniques to keep
the state private with the same efficiency benefits.
When presenting our approach, we compare facts about private inputs
that it reveals to those revealed from public loop bounds
(\cref{sec:privacy}).

vRAM~\cite{zhang2018vram} has achieved further efficiency with a
ingenious universal preprocessing that allows the parties to use a
smaller circuit tailored to verifying the specific program on the
chosen inputs.
Unfortunately, such tailored circuits can reveal significant
information about the input provided.
%
%
Our public-pc optimization (\cref{sec:publicpc}) attempts to balance
the gains of a tailored circuit and the privacy requirements of the
prover.

\section{Conclusion}
\label{sec:conclusion}
Due to a sustainted successes in the development of ZK protocols,
recent techniques have reached the cusp of proving knowledge of
realistic vulnerabilities and proving subtle exploits in low-level
code.
This paper describes how a host of core techniques from compiler
design---namely, conservative instruction profiling and
under-approximating information-flow tainting---can be implemented in
an optimizing proof-statement generator to produce proofs of
vulnerability in commodity software that can be triggered only be
using a considerable amount of time and space.

Our practical experience has produced a zero-knowledge proof of memory
unsafety in FFmpeg and a proof of leakakge in OpenSSL that directly
used the Heartbleed exploit as a witness and demonstrates that zero
knowledge proofs of vulnerability in critical application software are
now practical.


\section*{Availability and Ethical Considerations}
We are in process of open sourcing the implementation of \toolname{}
for publication and artifact evaluation.  \toolname{} aids in
responsible disclosure by producing zero-knowledge proofs of the
existence of vulnerabilities while keeping the vulnerabilities and
exploits secret.
All vulnerabilities used in our evaluation have been previously disclosed publicly, and fixes are widely deployed.
Thus, the work presented in this paper does not constitute an
unethical disclosure of potentially harmful information.

\section*{Acknowledgments}

This material is based upon work supported by the Defense Advanced Research
Projects Agency (DARPA) under Contract No.\ HR001120C0085. Any opinions,
findings and conclusions or recommendations expressed in this material are those
of the author(s) and do not necessarily reflect the views of the Defense
Advanced Research Projects Agency (DARPA). 
Approved for Public Release, Distribution Unlimited.

\bibliographystyle{plain}
\bibliography{cheesecloth,cryptobib/abbrev2,cryptobib/crypto_crossref}

\begin{thebibliography}{10}

\bibitem{cve-2013-0864}
{CVE-2013-0864}.
\newblock \url{https://cve.mitre.org/cgi-bin/cvename.cgi?name=CVE-2013-0864}.
\newblock Accessed: 2022-10-10.

\bibitem{ffmpeg}
{FFmpeg}.
\newblock \url{https://ffmpeg.org/}.
\newblock Accessed: 2022-09-01.

\bibitem{openssl}
{OpenSSL}: Cryptography and {SSL}/{TLS} toolkit.
\newblock \url{https://openssl.org/}.
\newblock Accessed: 2022-09-05.

\bibitem{picolibc}
{Picolibc}: {C} libraries for smaller embedded systems.
\newblock \url{https://keithp.com/picolibc/}.
\newblock Accessed: 2022-10-10.

\bibitem{heartbleed}
{The Heartbleed Bug}.
\newblock \url{https://heartbleed.com/}.
\newblock Accessed: 2022-09-05.

\bibitem{sieve-ir}
{zkInterface}: {SIEVE} intermediate representation {(IR)} proposal.
\newblock \url{https://hackmd.io/@danib31/BkP9HBp2L}.
\newblock Accessed: 2022-10-10.

\bibitem{aho2007compilers}
Alfred~V Aho, Monica~S Lam, Ravi Sethi, and Jeffrey~D Ullman.
\newblock {\em Compilers: principles, techniques, \& tools}.
\newblock Pearson Education India, 2007.

\bibitem{CCS:AHIV17}
Scott Ames, Carmit Hazay, Yuval Ishai, and Muthuramakrishnan
  Venkitasubramaniam.
\newblock Ligero: Lightweight sublinear arguments without a trusted setup.
\newblock In Bhavani~M. Thuraisingham, David Evans, Tal Malkin, and Dongyan Xu,
  editors, {\em ACM CCS 2017}, pages 2087--2104, Dallas, TX, USA,
  October~31~--~November~2, 2017. {ACM} Press.

\bibitem{C:BMRS21}
Carsten Baum, Alex~J. Malozemoff, Marc~B. Rosen, and Peter Scholl.
\newblock Mac'n'cheese: Zero-knowledge proofs for boolean and arithmetic
  circuits with nested disjunctions.
\newblock In Malkin and Peikert \cite{C21-4}, pages 92--122.

\bibitem{C:BCGTV13}
Eli {Ben-Sasson}, Alessandro Chiesa, Daniel Genkin, Eran Tromer, and Madars
  Virza.
\newblock {SNARKs} for {C}: Verifying program executions succinctly and in zero
  knowledge.
\newblock In Ran Canetti and Juan~A. Garay, editors, {\em CRYPTO~2013,
  Part~II}, volume 8043 of {\em {LNCS}}, pages 90--108, Santa Barbara, CA, USA,
  August~18--22, 2013. Springer, Heidelberg, Germany.

\bibitem{ben2013tinyram}
Eli Ben-Sasson, Alessandro Chiesa, Daniel Genkin, Eran Tromer, and Madars
  Virza.
\newblock {TinyRAM} architecture specification, v0.991.
\newblock \url{https://www.scipr-lab.org/doc/TinyRAM-spec-0.991.pdf}, 2013.

\bibitem{ben2014succinct}
Eli Ben-Sasson, Alessandro Chiesa, Eran Tromer, and Madars Virza.
\newblock Succinct non-interactive zero knowledge for a {von Neumann}
  architecture.
\newblock In {\em 23rd {USENIX} Security Symposium ({USENIX} Security 14)},
  pages 781--796, 2014.

\bibitem{benton2004simple}
Nick Benton.
\newblock Simple relational correctness proofs for static analyses and program
  transformations.
\newblock {\em ACM SIGPLAN Notices}, 39(1):14--25, 2004.

\bibitem{TCC:BHRRS20}
Alexander~R. Block, Justin Holmgren, Alon Rosen, Ron~D. Rothblum, and Pratik
  Soni.
\newblock Public-coin zero-knowledge arguments with (almost) minimal time and
  space overheads.
\newblock In Rafael Pass and Krzysztof Pietrzak, editors, {\em TCC~2020,
  Part~II}, volume 12551 of {\em {LNCS}}, pages 168--197, Durham, NC, USA,
  November~16--19, 2020. Springer, Heidelberg, Germany.

\bibitem{C:BHRRS21}
Alexander~R. Block, Justin Holmgren, Alon Rosen, Ron~D. Rothblum, and Pratik
  Soni.
\newblock Time- and space-efficient arguments from groups of unknown order.
\newblock In Malkin and Peikert \cite{C21-4}, pages 123--152.

\bibitem{AC:BCGJM18}
Jonathan Bootle, Andrea Cerulli, Jens Groth, Sune~K. Jakobsen, and Mary Maller.
\newblock Arya: Nearly linear-time zero-knowledge proofs for correct program
  execution.
\newblock In Thomas Peyrin and Steven Galbraith, editors, {\em ASIACRYPT~2018,
  Part~I}, volume 11272 of {\em {LNCS}}, pages 595--626, Brisbane, Queensland,
  Australia, December~2--6, 2018. Springer, Heidelberg, Germany.

\bibitem{braun2013pantry}
Benjamin Braun, Ariel~J Feldman, Zuocheng Ren, Srinath Setty, Andrew~J
  Blumberg, and Michael Walfish.
\newblock Verifying computations with state.
\newblock In {\em Proceedings of the Twenty-Fourth ACM Symposium on Operating
  Systems Principles}, pages 341--357, 2013.

\bibitem{clarkson2010hyperproperties}
Michael~R Clarkson and Fred~B Schneider.
\newblock Hyperproperties.
\newblock {\em Journal of Computer Security}, 18(6):1157--1210, 2010.

\bibitem{denning1976lattice}
Dorothy~E Denning.
\newblock A lattice model of secure information flow.
\newblock {\em Communications of the ACM}, 19(5):236--243, 1976.

\bibitem{enck2014taintdroid}
William Enck, Peter Gilbert, Seungyeop Han, Vasant Tendulkar, Byung-Gon Chun,
  Landon~P Cox, Jaeyeon Jung, Patrick McDaniel, and Anmol~N Sheth.
\newblock Taintdroid: an information-flow tracking system for realtime privacy
  monitoring on smartphones.
\newblock {\em ACM Transactions on Computer Systems (TOCS)}, 32(2):1--29, 2014.

\bibitem{fang2021zero}
Zhiyong Fang, David Darais, Joseph~P Near, and Yupeng Zhang.
\newblock Zero knowledge static program analysis.
\newblock In {\em Proceedings of the 2021 ACM SIGSAC Conference on Computer and
  Communications Security}, pages 2951--2967, 2021.

\bibitem{CCS:FKLOWW21}
Nicholas Franzese, Jonathan Katz, Steve Lu, Rafail Ostrovsky, Xiao Wang, and
  Chenkai Weng.
\newblock Constant-overhead zero-knowledge for {RAM} programs.
\newblock In Giovanni Vigna and Elaine Shi, editors, {\em ACM CCS 2021}, pages
  178--191, Virtual Event, Republic of Korea, November~15--19, 2021. {ACM}
  Press.

\bibitem{swanky}
{Galois, Inc.}
\newblock {swanky}: A suite of rust libraries for secure computation.
\newblock \url{https://github.com/GaloisInc/swanky}, 2019.

\bibitem{EC:GGPR13}
Rosario Gennaro, Craig Gentry, Bryan Parno, and Mariana Raykova.
\newblock Quadratic span programs and succinct {NIZKs} without {PCPs}.
\newblock In Thomas Johansson and Phong~Q. Nguyen, editors, {\em
  EUROCRYPT~2013}, volume 7881 of {\em {LNCS}}, pages 626--645, Athens, Greece,
  May~26--30, 2013. Springer, Heidelberg, Germany.

\bibitem{goguen1982security}
Joseph~A Goguen and Jos{\'e} Meseguer.
\newblock Security policies and security models.
\newblock In {\em 1982 IEEE Symposium on Security and Privacy}, pages 11--11.
  IEEE, 1982.

\bibitem{goldreich1991proofs}
Oded Goldreich, Silvio Micali, and Avi Wigderson.
\newblock Proofs that yield nothing but their validity or all languages in np
  have zero-knowledge proof systems.
\newblock {\em Journal of the ACM (JACM)}, 38(3):690--728, 1991.

\bibitem{green2022efficient}
Matthew Green, Mathias Hall-Andersen, Eric Hennenfent, Gabriel Kaptchuk,
  Benjamin Perez, and Gijs Van~Laer.
\newblock Efficient proofs of software exploitability for real-world
  processors.
\newblock {\em Cryptology ePrint Archive}, 2022.

\bibitem{EC:Groth16}
Jens Groth.
\newblock On the size of pairing-based non-interactive arguments.
\newblock In Marc Fischlin and Jean-S{\'{e}}bastien Coron, editors, {\em
  EUROCRYPT~2016, Part~II}, volume 9666 of {\em {LNCS}}, pages 305--326,
  Vienna, Austria, May~8--12, 2016. Springer, Heidelberg, Germany.

\bibitem{heath20202}
David Heath and Vladimir Kolesnikov.
\newblock A 2.1 khz zero-knowledge processor with bubbleram.
\newblock In {\em Proceedings of the 2020 ACM SIGSAC Conference on Computer and
  Communications Security}, pages 2055--2074, 2020.

\bibitem{heath2020stacked}
David Heath and Vladimir Kolesnikov.
\newblock Stacked garbling for disjunctive zero-knowledge proofs.
\newblock In {\em Annual International Conference on the Theory and
  Applications of Cryptographic Techniques}, pages 569--598. Springer, 2020.

\bibitem{SP:HYDK21}
David Heath, Yibin Yang, David Devecsery, and Vladimir Kolesnikov.
\newblock Zero knowledge for everything and everyone: Fast {ZK} processor with
  cached {ORAM} for {ANSI} {C} programs.
\newblock In {\em 2021 {IEEE} Symposium on Security and Privacy}, pages
  1538--1556, San Francisco, CA, USA, May~24--27, 2021. {IEEE} Computer Society
  Press.

\bibitem{C:HuMohRos15}
Zhangxiang Hu, Payman Mohassel, and Mike Rosulek.
\newblock Efficient zero-knowledge proofs of non-algebraic statements with
  sublinear amortized cost.
\newblock In Rosario Gennaro and Matthew J.~B. Robshaw, editors, {\em
  CRYPTO~2015, Part~II}, volume 9216 of {\em {LNCS}}, pages 150--169, Santa
  Barbara, CA, USA, August~16--20, 2015. Springer, Heidelberg, Germany.

\bibitem{ishai2007zero}
Yuval Ishai, Eyal Kushilevitz, Rafail Ostrovsky, and Amit Sahai.
\newblock Zero-knowledge from secure multiparty computation.
\newblock In {\em Proceedings of the thirty-ninth annual ACM symposium on
  Theory of computing}, pages 21--30, 2007.

\bibitem{grit}
{Jasper Vijn}.
\newblock {GRIT}: Gba raster image transmogrifier.
\newblock \url{https://github.com/devkitPro/grit}, 2022.

\bibitem{jones1993partial}
Neil~D Jones, Carsten~K Gomard, and Peter Sestoft.
\newblock {\em Partial evaluation and automatic program generation}.
\newblock Peter Sestoft, 1993.

\bibitem{katz2018improved}
Jonathan Katz, Vladimir Kolesnikov, and Xiao Wang.
\newblock Improved non-interactive zero knowledge with applications to
  post-quantum signatures.
\newblock In {\em Proceedings of the 2018 ACM SIGSAC Conference on Computer and
  Communications Security}, pages 525--537, 2018.

\bibitem{C21-4}
Tal Malkin and Chris Peikert, editors.
\newblock {\em CRYPTO~2021, Part~IV}, volume 12828 of {\em {LNCS}}, Virtual
  Event, August~16--20, 2021. Springer, Heidelberg, Germany.

\bibitem{EC:MohRosSca17}
Payman Mohassel, Mike Rosulek, and Alessandra Scafuro.
\newblock Sublinear zero-knowledge arguments for {RAM} programs.
\newblock In Jean-S{\'{e}}bastien Coron and Jesper~Buus Nielsen, editors, {\em
  EUROCRYPT~2017, Part~I}, volume 10210 of {\em {LNCS}}, pages 501--531, Paris,
  France, April~30~--~May~4, 2017. Springer, Heidelberg, Germany.

\bibitem{myers1999jflow}
Andrew~C Myers.
\newblock Jflow: Practical mostly-static information flow control.
\newblock In {\em Proceedings of the 26th ACM SIGPLAN-SIGACT symposium on
  Principles of programming languages}, pages 228--241, 1999.

\bibitem{myers1997decentralized}
Andrew~C Myers and Barbara Liskov.
\newblock A decentralized model for information flow control.
\newblock {\em ACM SIGOPS Operating Systems Review}, 31(5):129--142, 1997.

\bibitem{10.1145/1250734.1250746}
Nicholas Nethercote and Julian Seward.
\newblock Valgrind: A framework for heavyweight dynamic binary instrumentation.
\newblock In {\em Proceedings of the 28th ACM SIGPLAN Conference on Programming
  Language Design and Implementation}, PLDI '07, page 89–100, New York, NY,
  USA, 2007. Association for Computing Machinery.

\bibitem{10.1145/3290388}
James Parker, Niki Vazou, and Michael Hicks.
\newblock Lweb: Information flow security for multi-tier web applications.
\newblock {\em Proc. ACM Program. Lang.}, 3(POPL), jan 2019.

\bibitem{SP:PHGR13}
Bryan Parno, Jon Howell, Craig Gentry, and Mariana Raykova.
\newblock Pinocchio: Nearly practical verifiable computation.
\newblock In {\em 2013 {IEEE} Symposium on Security and Privacy}, pages
  238--252, Berkeley, CA, USA, May~19--22, 2013. {IEEE} Computer Society Press.

\bibitem{reynolds2002separation}
John~C Reynolds.
\newblock Separation logic: A logic for shared mutable data structures.
\newblock In {\em Proceedings 17th Annual IEEE Symposium on Logic in Computer
  Science}, pages 55--74. IEEE, 2002.

\bibitem{sabelfeld2003language}
Andrei Sabelfeld and Andrew~C Myers.
\newblock Language-based information-flow security.
\newblock {\em IEEE Journal on selected areas in communications}, 21(1):5--19,
  2003.

\bibitem{sagiv2002parametric}
Mooly Sagiv, Thomas Reps, and Reinhard Wilhelm.
\newblock Parametric shape analysis via 3-valued logic.
\newblock {\em ACM Transactions on Programming Languages and Systems (TOPLAS)},
  24(3):217--298, 2002.

\bibitem{shapiro1997fast}
Marc Shapiro and Susan Horwitz.
\newblock Fast and accurate flow-insensitive points-to analysis.
\newblock In {\em Proceedings of the 24th ACM SIGPLAN-SIGACT symposium on
  Principles of programming languages}, pages 1--14, 1997.

\bibitem{steensgaard1996points}
Bjarne Steensgaard.
\newblock Points-to analysis in almost linear time.
\newblock In {\em Proceedings of the 23rd ACM SIGPLAN-SIGACT symposium on
  Principles of programming languages}, pages 32--41, 1996.

\bibitem{stefan2011flexible}
Deian Stefan, Alejandro Russo, John~C Mitchell, and David Mazi{\`e}res.
\newblock Flexible dynamic information flow control in haskell.
\newblock In {\em Proceedings of the 4th ACM Symposium on Haskell}, pages
  95--106, 2011.

\bibitem{suh2004secure}
G~Edward Suh, Jae~W Lee, David Zhang, and Srinivas Devadas.
\newblock Secure program execution via dynamic information flow tracking.
\newblock {\em ACM Sigplan Notices}, 39(11):85--96, 2004.

\bibitem{NDSS:WSRBW15}
Riad~S. Wahby, Srinath T.~V. Setty, Zuocheng Ren, Andrew~J. Blumberg, and
  Michael Walfish.
\newblock Efficient {RAM} and control flow in verifiable outsourced
  computation.
\newblock In {\em NDSS~2015}, San Diego, CA, USA, February~8--11, 2015. The
  Internet Society.

\bibitem{wahby2015buffet}
Riad~S Wahby, Srinath~TV Setty, Zuocheng Ren, Andrew~J Blumberg, and Michael
  Walfish.
\newblock Efficient {RAM} and control flow in verifiable outsourced
  computation.
\newblock In {\em NDSS}, 2015.

\bibitem{zhang2018vram}
Yupeng Zhang, Daniel Genkin, Jonathan Katz, Dimitrios Papadopoulos, and
  Charalampos Papamanthou.
\newblock {vRAM}: Faster verifiable ram with program-independent preprocessing.
\newblock In {\em 2018 IEEE Symposium on Security and Privacy (SP)}, pages
  908--925. IEEE, 2018.

\end{thebibliography}

\appendix


\end{document}